# Using a coupled optical and electrical monitoring method to follow the R-HiPIMS TiO$_2$ deposition process drifts.

D. Boivin[1], R. Jean-Marie-Désirée[1], A. Najah[1], S. Cuynet[1] and L. de Poucques[1]

[1]*Université de Lorraine, CNRS, IJL, F-54000 Nancy, France*
*e-mail : ronny.jean-marie-desiree@univ-lorraine.fr*

**Abstract:**

In this work, coupled optical and electrical discharge measurements have been implemented to investigate the plasma state of a reactive HiPIMS TiO$_2$ deposition process running at a fixed duty cycle of 2% and at a repetition rate of 1 kHz. Investigations focus on both the effect of the erosion target and substrate-holder temperature in an Ar/O$_2$ gas mixture at fixed working pressure. First, as the racetrack shape evolves with the use of the target, the deposition rate is modified, in the same way as the emission intensity measured. Second, with the heater set at 400 °C in pure Ar, the coating appears thinner, and the optical emission spectroscopy measurement reveals the presence of oxygen atoms. Thus, the results from the coupled optical and electrical measurements are sensitive enough to track potential drift of the process.

## 1. Introduction:

Titanium oxide materials, due to its various chemical and physical properties, such as high dielectric constant [1], high refraction index [2], good thermal and chemical stability [3], catalytic photoactivity [4], and biocompatibility [5], are widely used in various industrial applications. These latter include optical and electronic devices [6-9], gas detection [10-12], water purification [13-15], and solar cells [16-18]. Titanium oxide thin films can be deposited using conventional DC or RF magnetron sputtering [19-22], as well as high-power impulse magnetron sputtering (HiPIMS) [23-25]. This latter is a type of ionized physical vapor deposition (IPVD) process [26-28] that utilizes a magnetron cathode. However, due to the magnetron architecture, the target surface is not homogeneously sputtered. Hence, a "racetrack" is formed as the target is consumed [29]. HiPIMS can achieve a high ionization degree of the sputtered atoms, due to the high applied voltage for a short lapse of time. Then, the power of the discharge is about 100 times higher than a conventional magnetron one. This high ionization rate is a key feature for a better conformal coating to be deposited on complex-shaped substrates through the application of a bias voltage to the substrate [30].

Moreover the high ion bombardment of the growing film can also be a significant parameter that influences the quality of the coatings [31-33]. HiPIMS can also be operated in reactive mode (R-HiPIMS) when using reactive gas such as $O_2$ or $N_2$ mixed with a non-reactive one to deposit coatings such as oxides or nitrides. Nevertheless, the poisoning of the target induced by the reactive gas makes the control of the process more complex since a transition occurs, from a metallic to a compound system. Since the nature of the initial sputtered material is altered, parameters associated to the discharge (*e.g.,* voltage [34-35] and current [36]) and the ones to the sputtering (*e.g.* sputtering yield, deposition rate) change. In addition, these parameters exhibit a hysteresis that has been extensively studied [37-38]. Henceforth, the coating industry requires an easy and reliable method to track the evolution of the deposition over time. To address this need, many options have been studied: using a feedback control loop on the reactive gas flow [39-40], regulating the pulse frequency [41-44], monitoring the intensity of an emission line (plasma emission monitoring) [45-48], or using ratio between a line intensity measured by optical emission spectroscopy (OES) and the current density [49]. One limitation when using a feedback control loop on the reactive gas flow is the response time delay needed to stabilize the plasma state. Other methods involve tuning the magnetic field strength surrounding the target [50-51], which are not practical for industrial installations.

In a previous work, we demonstrated the correlation between the emission line intensity (noted *TA_OES*) divided by the integrated current (noted $I_{int}$) data with layer thicknesses as a function of the proportion of $O_2$ added in the Ar gas and the pulse width [29]. In this work, we aim to improve the accuracy of this correlation by taking into account the process drifts (consumption of the target, reactor wall pollution). We will then use this to optimize the deposition process in order to achieve consistent coatings. From this, it was established that plasma diagnostics could be used effectively to control the deposition rate and precisely manage the transition between metal and compound sputtering modes. Using these to calibrate a power feedback control loop could then provide a better response time compared to gas feedback control loop.

In this study, our method is proposed to enforce the feedback control loop on electrical discharge parameters (as suggested in [41-44]) with coupled electrical and optical emission spectroscopy measurements (as described in our previous work [29]) to follow the R-HiPIMS deposition process drifts of $TiO_2$ as an example. These measurements are investigated for two series of experiments. The first one consists of studying the influence of the racetrack formation along the deposition process and the correlations that we can draw from this, in relation to our monitoring method; the second one deals with the factual possibility of working in the metallic sputtering mode at 0% $O_2$ or at very low % $O_2$ in the gas mixture, with or without additional heating temperature of the substrate-holder for both cases.

## 2. Experimental set-up

In this study, the plasma reactor is a stainless-steel chamber with a main cylindrical body of 35 cm in diameter and 30 cm in height. A water-cooled circular planar balanced magnetron cathode, of which magnetic characteristic had been described in previous article [29], is set in the chamber.

The target used is a 2" diameter and 3 mm thickness of pure titanium (99.99%). All substrates are placed 11 cm from the target surface, on a grounded substrate-holder. The chamber is pumped by a primary pump (ACP 15 Adixen) and a turbomolecular pump (ATP 400 Adixen) down to a residual pressure of $1.2\times10^{-4}$ Pa before each deposition. The pressure inside the chamber is measured by a AHC 2010 Adixen gauge, calibrated by a capacitive one. The gas flow injection is located at 15 cm from the target centre and parallel to the target surface. The total gas flow (Ar + $O_2$) is maintained to 19 sccm (standard cubic centimetre per minute) using two mass flowmeters (20 sccm and 3 sccm full scale, respectively). This is achieved by decreasing the Ar gas flow when the $O_2$ one is increased. The injection of $O_2$, up to 2 % of the total gas flow (the maximal range studied), does not lead to a variation of the working pressure. Thanks to the throttle valve, the working pressure is set to 0.7 Pa. The % $O_2$ refers to the percentage of $O_2$-to-Ar gas flow ratio.

The cathode is connected to a high voltage generator (MELEC SIPP 2000). The high voltage pulse waveform is driven by an arbitrary function generator (Tektronix AFG 3022 C) The mean power $<P>_T$ (eq. 1) during the HiPIMS period $T$ and the voltage pulse frequency are set at $\approx$ 45 W and 1 kHz, respectively. During this study, the pulse width setting $T_d$ is maintained at 20 µs, which determines the discharge duration. Therefore, the mean power during this discharge duration is around 2250 W according to eq. 2.

$$\langle P \rangle_T = \frac{1}{T}\int_0^T P(t)dt \qquad (1)$$

$$\langle P \rangle_{T_d} = \frac{1}{T_d}\int_0^{T_d} P(t)dt \qquad (2)$$

To measure the voltage-current time characteristics of the discharge, a voltage probe from CalTest Electronics CT4028 with a bandwidth from DC to 220 MHz and a current probe from MagneLab CT0.1-B, bandwidth up to 50 MHz were used with via an oscilloscope (Lecroy Wave runner 104Xi 1GHz). For the purpose of this study, the current is integrated on the duration of the pulse current, as described by figure 5 in our previous work [29], noted $I_{int}$.

Time-averaged optical emission spectroscopy measurements ($TA\_OES$) were carried out with a Jobin Yvon TRIAX 550 spectrometer, equipped with a 1800 tr.mm$^{-1}$ grating, a 100 µm slit and an intensified charge-coupled device (ICCD). The light emitted by the plasma is collected through a cylindrical collimator (9 cm length over 1 cm diameter) placed parallel to the target surface. The titanium emission line intensity (at 365.35 nm) is then divided by $I_{int}$, to take into account the effect of any variation in electron population, assuming the excitation process is mainly due by electron collisions. More details about the $TA\_OES/I_{int}$ measurements can be found in our previous work [29]. The oxygen emission line has been resolved allowing to consider only the emission line intensity at 777.19 nm.

The process of depositing thin films on flat silicon (Si) wafers involved using a heating substrate-holder from UHV designer. The substrate-holder and samples were heated by IR radiation emitted from a heater made of sSiC. The temperature was set at 400 °C and the deposition time was fixed at 4 hours, irrespective of the magnetron sputtering conditions.

All substrates were cleaned in the same way, using first an ultrasonic bath of acetone and then ethanol. The wafer was then rinsed and dried using deionized water and nitrogen gas, respectively. The substrate was hooked to the grounded substrate-holder using metallic clamps combined with a Si piece of 60×10 cm² as a mask. The film thickness was measured by profilometry (Bruker Dektak XT) equipped with a diamond tip of 2 µm radius, after removing the Si mask. As the thickness values obtained are in good accordance with cross-sectional imaging using scanning electron microscopy (SEM Gemini Zeiss), only the profilometry measurements are discussed. Note that the purpose of this study is only to investigate the feasibility of using the presented monitoring method to detect process drifts during deposition, and not to study the quality of the deposits themselves.

To measure the erosion depth of the target racetrack, a palmer micrometre was used. This measurement was carried out on four distinct locations, to verify the homogeneity of the racetrack.

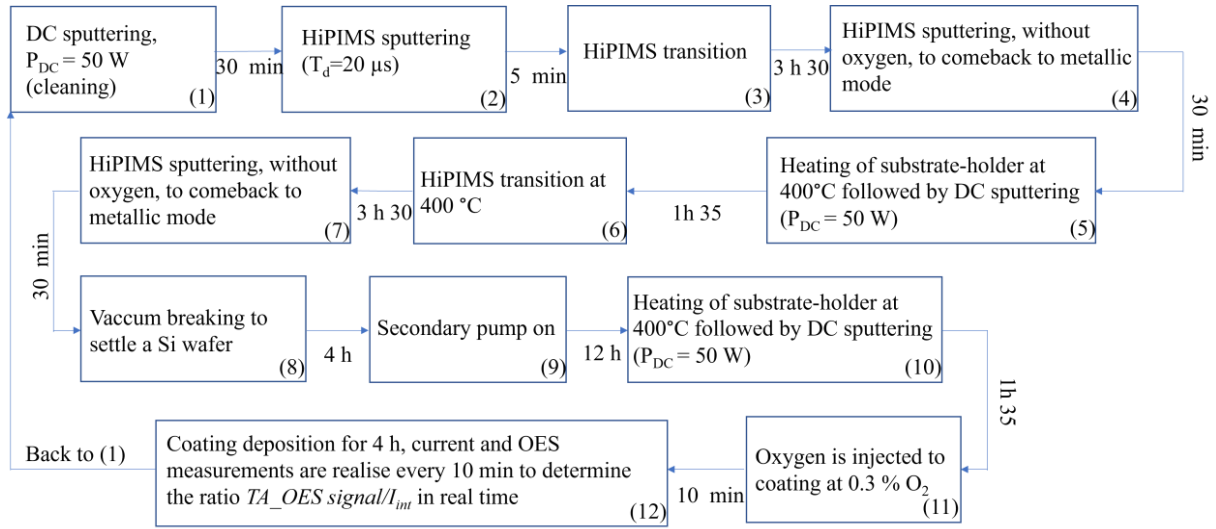

*Figure 1:* Synoptic diagram of the experimental method to study the evolution of the $TA\_OES/I_{int}$ ratio and coating thickness with the erosion target.

To strictly track the deposition and plasma characteristic throughout the sputtering of the target, an experimental protocol (diagrammed in figure 1) was developed as follows:

(1) Sputtering of the target, using DC mode with $P_{DC}$ = 50 W and under argon atmosphere at 0.7 Pa. This ensures to clean and condition the racetrack surface polluted when venting, (30 min).

(2) Switching with HiPIMS mode, with a discharge time of 20 µs to stabilize the process, *i.e.* when the discharge voltage does not vary significantly anymore, (5 min).

(3) OES study of the transition, from metallic to fully oxidized regime, in HiPIMS mode, without additional heating on the substrate-holder, (3h30).

(4) Sputtering of the target under the same conditions as the previous point, without oxygen, to return to the metallic regime, (about 30 min).

(5) Heating the substrate-holder during 1 h to reach 400 °C, followed by the cleaning step (1) and the stabilization step (2), (1h35 in total).

(6) OES study of the transition, from metallic to fully oxidized regime, in HiPIMS mode, at 400 °C, (3h30).

(7) Sputtering of the target under the same conditions as the previous point, without injected oxygen, to return to the metallic regime, heater turned off, (about 30 min).

(8) Cooling and venting of the chamber to install a silicon wafer (about 4 h).

(9) Pumping during 12h to reach a vacuum of about $10^{-4}$ Pa, (about 12h).

(10) Heating the substrate-holder for 1 h to reach 400 °C, followed by the cleaning step (1) and the stabilization step (2), (1h35 in total).

(11) Only for the 0.3 %O$_2$ deposition, injection of the gas mixture to use. A 10-minutes wait ensures a stable process with a well-mixed atmosphere.

(12) 4h deposition for a given plasma condition during which electrical and optical (Ti emission line at 365.35 nm) measurements are done every 10 min to determine the $TA\_OES/I_{int}$ ratio (4h).

The sample is then remove from the chamber and the process is repeat again from (1) to (12).

**Table 1:** Overview of the experimental conditions of the studied coatings.

| Coating number | 1 | 2 | 3 | 4 | 5 | 6 | 7 | 8 | 9 | 10 | 11 | 12 | 13 |
|---|---|---|---|---|---|---|---|---|---|---|---|---|---|
| % O$_2$ | 0.3 | 0.3 | 0 | 0 | 0 | 0 | 0 | 0.3 | 0.3 | 0 | 0 | 0.3 | 0 |
| Heating (°C) | 400 | 400 | 400 | 400 | 400 | 400 | 400 | 400 | 400 | 0 | 0 | 400 | 400 |

Even if all the transitions announced by this procedure (with or without additional heating) were systematically measured, only a few amounts of these measurements have been selected, those allowing the discussion. In total, 13 coatings have been done under different conditions listed in the table 1, knowing that the working pressure was set to 0.7 Pa, the frequency to 1 kHz, the discharge pulse width to 20 µs and the mean power to 45 W keeping the same target.

### 3. Results

#### 3.1. Influence of the target erosion state

To study the influence of the racetrack formation during the deposition process, optical emission spectroscopy (TA_OES) coupled with electric measurements [29] was carried out, tracking the time evolution of Ti emission line at 365.35 nm as a function of O$_2$ admixture named % O$_2$.

The first transition (figure 2(a)) was done right after the target conditioning, in DC discharge. The second (figure 2(b)) was obtained with a target whose maximum depth of the racetrack is about 0.75 mm. The third (figure 2(c)) and the last (figure 2(d)) were carried out for a racetrack depth of about 1.25 mm and 1.65 mm, respectively. Regardless the depth of the racetrack, three modes can be observed. A metallic mode for [O$_2$] = 0 %, a fully oxidized mode starting over 1 % and an intermediate one localized in-between. These modes have been described and detailed in previous work [29].

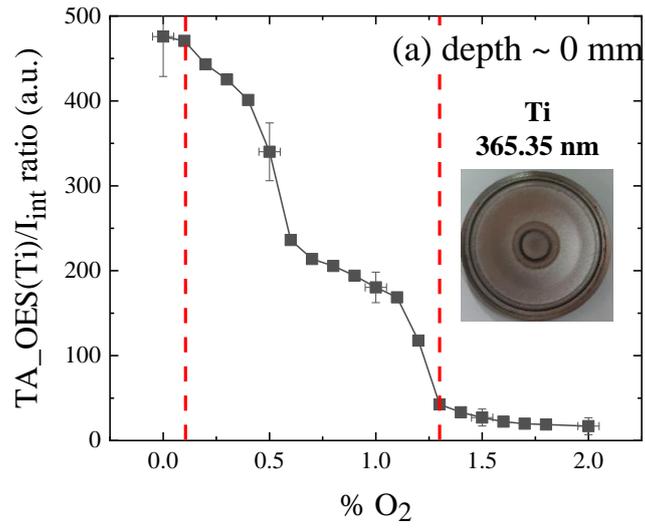

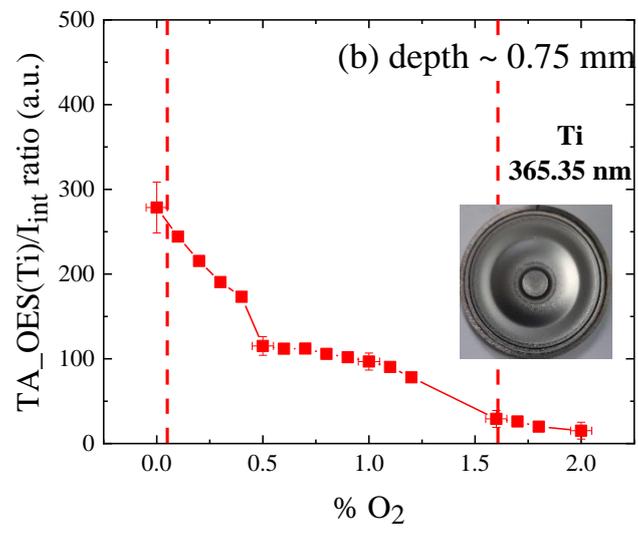

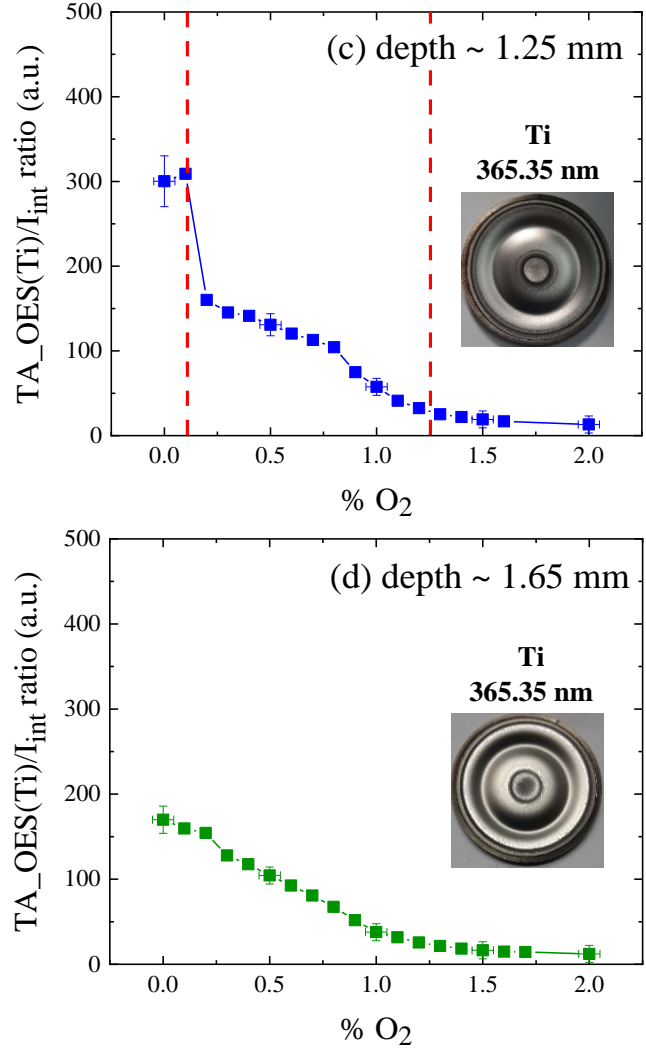

*Figure 2:* Transition, described with the TA_OES(Ti)/$I_{int}$ ratio, measured as a function of % $O_2$ in the gas mixture at different depth at the centre of the racetrack. (a): new target with 2h of DC sputtering; (b): depth at the centre of the racetrack 0.75 mm; (c): depth at the centre of the racetrack 1.25 mm; (d) : depth at the centre of the racetrack 1.65 mm. Append, the erosion target photographs corresponding to different depths at the centre racetrack. The pressure is 0.7 Pa, the HiPIMS period is 1 ms, $T_d$ = 20 µs and the mean power is 45 W

Overall, the transition pattern is significantly modified according to the racetrack depth, especially for low % $O_2$ (≤ 1 %) for which the *TA_SOE/$I_{int}$* ratio ratio drops with the erosion. Due to this drop, the intermediate mode is no longer distinctly visible on the transition at 1.65 mm. However, it is still present, otherwise the transition from the metallic to fully oxidized mode would be much steeper, as in the case of the transition between 1 and 1.2 % $O_2$ in DC discharge (*cf.* figure 8f from [29]).

More precisely, the phenomenon related to the target erosion is most likely due to the variation of the racetrack shape (depicted in figure 2, insert). Based on Rossnagel *et al* model [52] considering a cone-shaped angular distribution of the sputtered atoms, each sputtered atom has an axial (perpendicular to the target) and a radial (parallel to the target) velocity components. For a quasi-new target (figure 2(a)), the atoms are mainly ejected perpendicularly to the target surface, hence toward the substrate, because the erosion profile is still very close to the starting surface, *i.e.* the new target surface [52, 54]. When the racetrack is dug deep enough through sputtering, it results in some of the sputtered atoms being ejected toward the walls rather than toward the substrate. These atoms, carrying a non-negligible radial velocity, would therefore no longer pass through the densest plasma, which is centred above the racetrack. Consequently, they are less and less excited atoms that could explain the decrease of *TA_OES*/$I_{int}$ ratio as a function of the racetrack depth (figure 2). This could have been confirmed by particle simulations testing the effect of the target erosion on the particle transport in HiPIMS, which is not the main scope of this study.

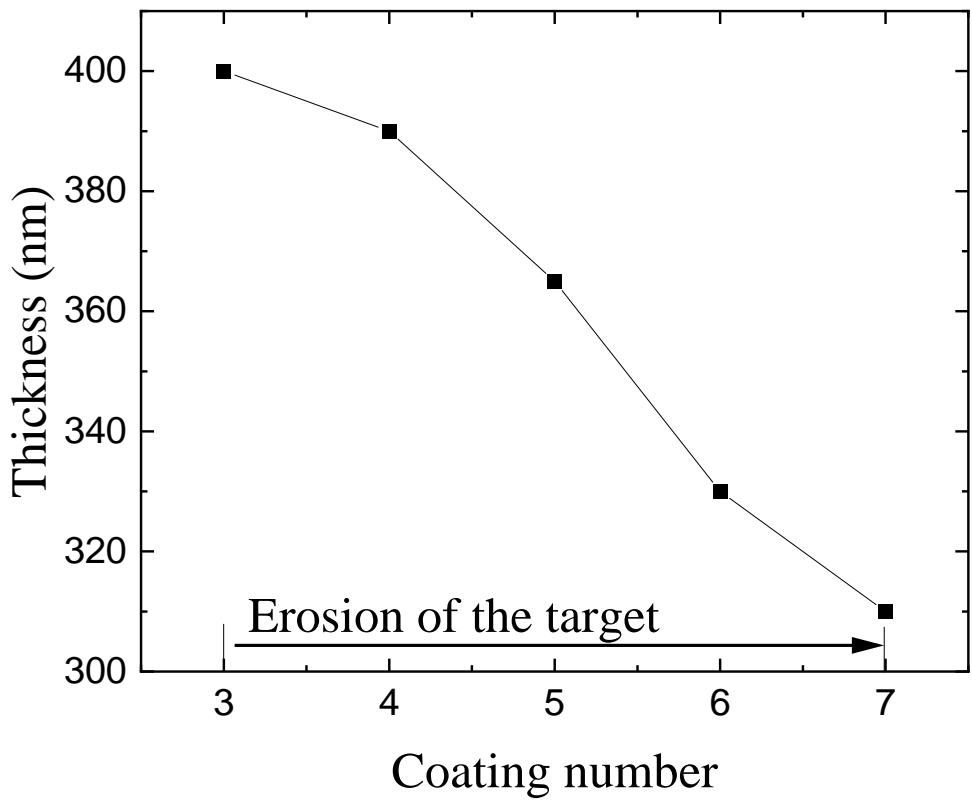

*Figure 3: Thickness evolution during erosion of the target. The pressure is 0.7 Pa, the deposition time is 4 hours, the HiPIMS period is 1 ms, in pure Ar gas, $T_d$ = 20 µs and the mean power is 45 W*

To close this section, a group of 5 coatings (numbered 4 – 8, *cf.* table 1) has been done using the same target and following the diagram from figure 1, to highlight the effect of target consumption on the coating thickness. Knowing that the sputtering process parameters (deposition time, pressure, etc.) remain the same for the samples presented in the figure 3, the coating thickness decreased from 400 nm to 310 nm. This tendency, in perspective with the decrease of $TA\_OES/I_{int}$ ratio over the racetrack depth (figure 2), is consistent with the description of the distribution of ejected atoms by Rossnagel *et al* [52] discussed above. A more recent work reported by Ganeva *et al.* [53] study the evolution of copper cluster intensity and size distribution produced by copper target sputtering and a gas aggregation source with the target lifetime. This study shows that the formation of these clusters depends on the target erosion which further supports the observed dependency between the coating thickness and the target erosion. Another hypothesis to explain the evolution of the thickness drawn in figure 3 could be the increase of the film density due to greater ion density. To discuss the latter, the influence of the discharge duration has been studied and published in our previous article, section 3.2. [29]. The increase of the pulse width ($T_d$) leads to an increase of the plasma power ($<P>_{Tint}$) which theoretically enhance the amount of titanium ions. However, the densification effect seems to be invalidated by figure 7 of our previous article [29], since $TA\_OES/I_{int}$ ratio of titanium neutral atoms is correlated with the layer thickness. Indeed, the sputtered species are increasingly ejected towards the walls as the racetrack becomes deeper. Therefore, the species densities that contribute to the coating become increasingly low.

### 3.2. Influence of the substrate-holder heating

To study the heating effect in depth, four coatings were carried out. The key conditions are listed in table 1, *ceteris paribus*. Note that the substrate-holder is pre-heated during one hour before the deposition process to reach 400 °C. TA-OES measurements were carried out during the deposition process and two transitions from metallic to fully oxidized mode, heated and unheated in accordance with the method described in section 2, figure 1. Concerning thickness measurement, summarized in table 2, it falls when an additional heating is used during the process. Taking the samples coated with no oxygen injected, the unheated one is around 1.3 times thicker than the heated one.

*Table 2: Experimental conditions and corresponding thickness for coatings heated at 400°C and without heating.*

| Coating number | 9 | 10 | 11 | 12 |
|---|---|---|---|---|
| % O$_2$ | 0.3 | 0 | 0 | 0.3 |

| Heating (°C) | 400 | 400 | 0 | 0 |
| Thickness (nm) | 92 | 240 | 315 | 150 |

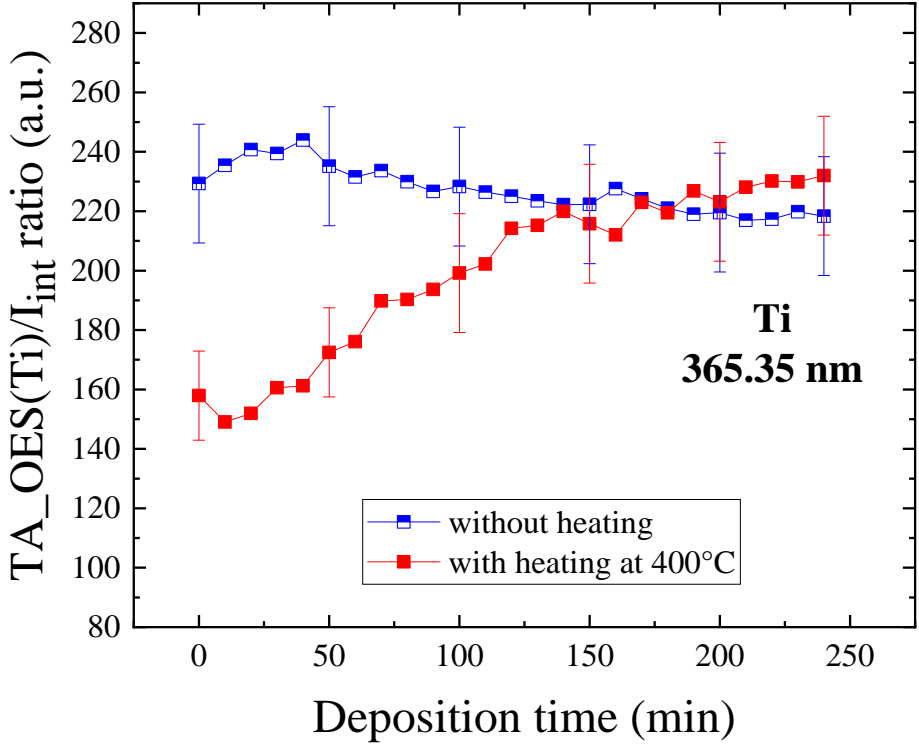

***Figure 4:*** *Evolution of the TA_OES (Ti)/I$_{int}$ ratio during the deposition, with heating at 400 °C, coating number 10, (red curve) and without heating, coating number 11, (blue curve). The pressure is 0.7 Pa, the deposition time is 4 hours, the HiPIMS period is 1 ms, in pure Ar gas, T$_d$ = 20 μs and the mean power is 45 W*

Figure 4 shows the evolution of *TA_OES/I$_{int}$* ratio done during the deposition process of the heated sample (red curve) and the unheated one (blue curve). Without additional heating, the *TA_OES(Ti)/I$_{int}$* ratio slightly varies of about 10 %, considering the global maximum and minimum values. More precisely, it decreases of about 5% between the beginning of the process (0 min) and the end (240 min). This tendency is consistent with the expected one, *i.e.* an almost constant ratio, even slightly decreasing due to the target sputtering procedure discussed in the previous section. On the contrary, the ratio measured with heating increases by about 70% for the first three hours before reaching a quasi-constant evolution in the last hour. Taking the uncertainty into account, the ratio tendency in the last hour is close to the one obtained during the unheated process.

To explain this phenomenon occurring with the heating, three hypotheses have been considered: a "volume-plasma" effect, a "surface-sputtering" effect, or both.

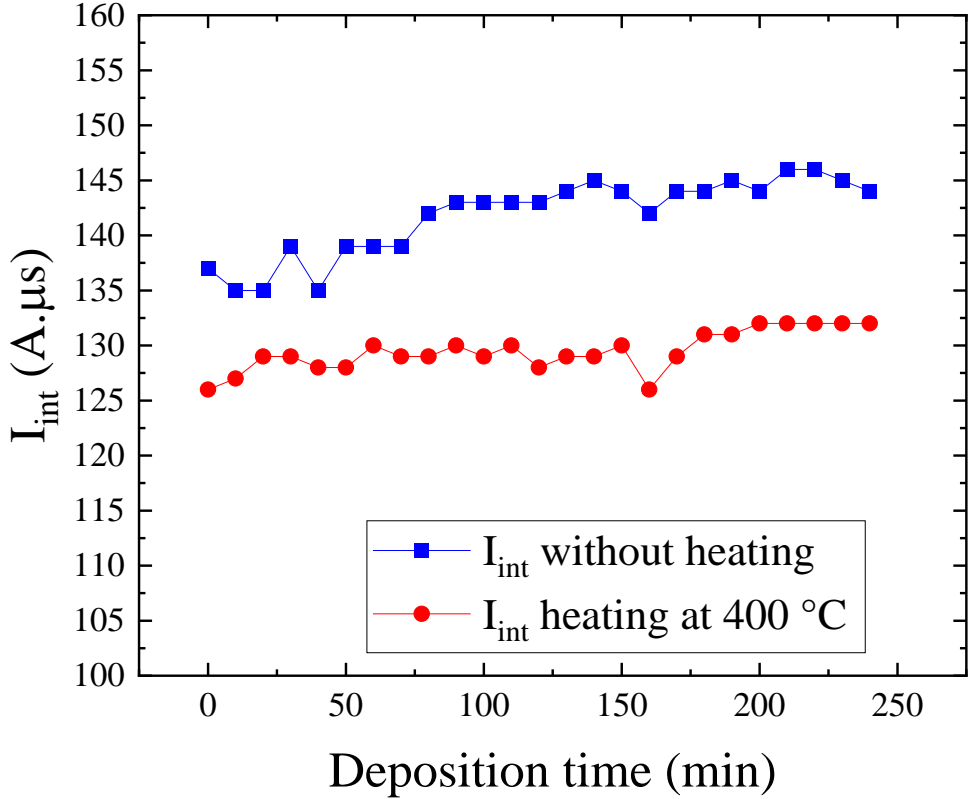

*Figure 5:* Evolution of the integrated (a) current $I_{int}$ and (b) voltage $V_{int}$ during the deposition, with heating at 400 °C, coating number 10, (red curve) and without heating, coating number 11, (blue curve). The pressure is 0.7 Pa, the deposition time is 4 hours, the HiPIMS period is 1 ms, in pure Ar gas, $T_d$= 20 µs and the mean power is 45 W

Concerning the first one, figure 5 presents the investigation of the electrical discharge parameter, *i.e.* integrated current $I_{int}$ depicts a slight variation (5 % at most) over the heated and unheated sputtering process. As the plasma is quite stable, it does not seem to be the main reason of $TA\_OES/I_{int}$ ratio variation over time and coating thickness (between heated and unheated coatings). In other words, the kinetic phenomenon described by the "heated" $TA\_OES(Ti)/I_{int}$ ratio over time (from figure 4) does not match the one described by the variation of the current on the sputtering yield. Therefore, this $TA\_OES/I_{int}$ ratio increasing over 4 hours may be induced by a slow-kinetic process, such as a drift phenomenon from surfaces. Heating the substrate-holder at 400 °C could be enough to lead to desorption of the reactor internal surfaces in our conditions. Moreover, the quantity of gas released is obviously relatively small since we did not observe any variation of the working pressure measured at 0.7 Pa, knowing that pressure control valve is set before heating the sample holder. To test this desorption hypothesis, a heating experiment in secondary vacuum was carried out for a whole day. Starting from a secondary vacuum, *i.e.* $3.99 \times 10^{-4}$ Pa before heating, the pressure increases slowly to a maximum value of 66.5 × 10⁻⁴ Pa after 1 hour, before decreasing slowly. This behaviour is attributed to a thermal wall desorption process which slightly enriches the volume with desorbed species containing oxygen (*e.g.* $H_2O$, $O_2$). Then, the initial vacuum before heating could not be reached again at the end of the day when heating is used.

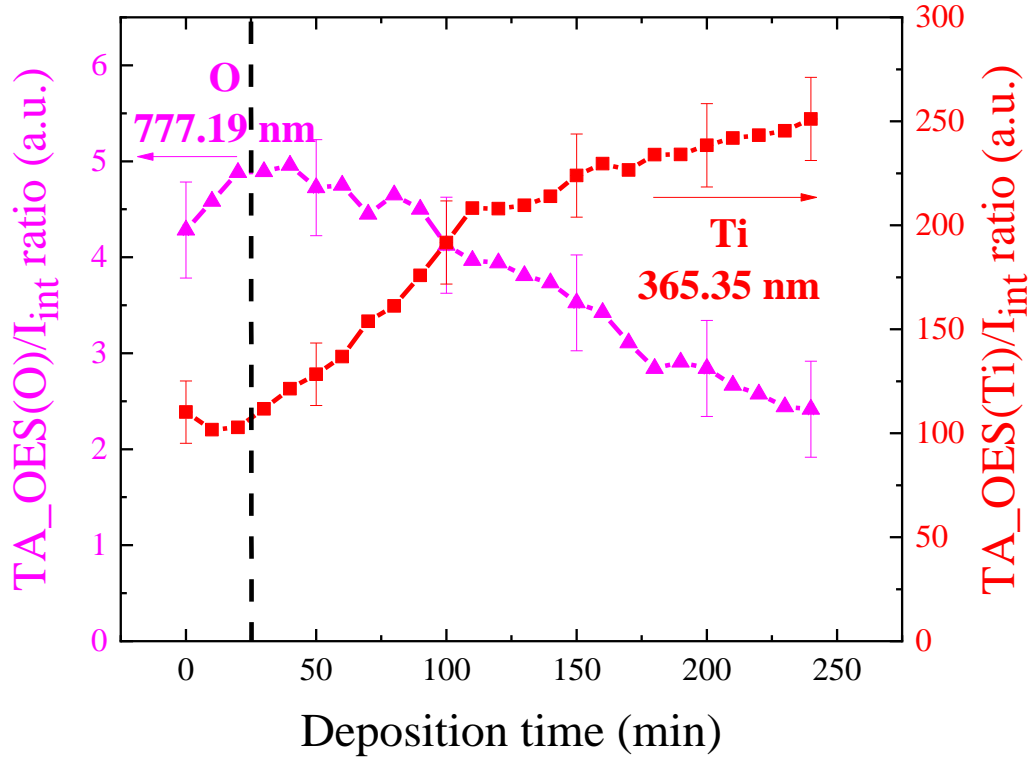

*Figure 6: Evolution of the TA_OES(O)/I$_{int}$ ratio and TA_OES(Ti)/I$_{int}$ ratio during the deposition with heating at 400 °C, coating number 12. The pressure is 0.7 Pa, the deposition time is 4 hours, the HiPIMS period is 1 ms, in pure Ar gas, $T_d$ = 20 µs and the mean power is 45 W*

Figure 6 compares the *TA_OES(O)/I$_{int}$* ratio, relative to the atomic oxygen line at 777.19 nm, with the *TA_OES(Ti)/I$_{int}$* ratio, relative to the atomic titanium line at 365.35 nm, measured every 10 min during the coating deposition with heating (coating number 12). From the beginning of the process, the *TA_OES(O)/I$_{int}$* ratio increases, reaching a maximum value at 25 min corresponding to 2h10 after turning on the heater, while no oxygen has been injected (pure argon gas is injected). When this heating-induced desorption seems maximal at 25 min, the *TA_OES(Ti)/I$_{int}$* ratio is minimal. Then, the *TA_OES(O)/I$_{int}$* ratio dwindles till the end of the 4-hour process, in contrast to *TA_OES(Ti)/I$_{int}$* ratio which increases. This opposite evolution of both ratios indicates a volume enrichment of titanium atoms and a depletion of oxygen atoms as the sputtering lasts. Since the "volume-plasma" effect hypothesis has been falsified with figure 5, the latter is due to a changing surface condition of the target, from partially oxidized by the outgassing oxygen-containing molecule to metallic. Indeed, as the target surface becomes less and less oxidized and poisoned, the sputtering yield increases as titanium metallic bonds is weaker ($E_{Ti-Ti} \approx 1.46$ eV $< E_{Ti-O} \approx 6.86$ eV, assuming standard temperature and pressure conditions [55]). Therefore, the observed phenomena depicted by figure 4 is a "surface-sputtering" effect where the increase of $TA\_OES(Ti)/I_{int}$ ratio with heating (red curve) reflects the target surface condition turning from partially oxidized to metallic on the first three hours.

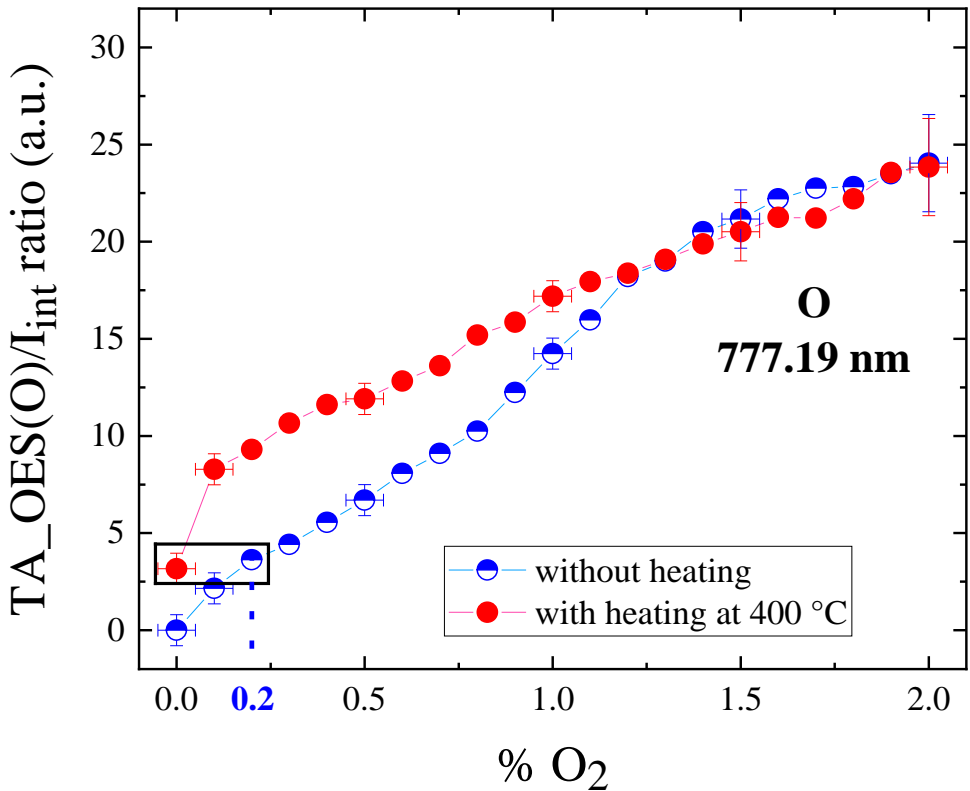

*Figure 7:* Evolution of $TA\_OES(O)/I_{int}$ ratio as a function of % $O_2$ in the gas mixture, with heating at 400 °C and without heating. The pressure is 0.7 Pa, the HiPIMS period is 1 ms, $T_d$ = 20 µs and the mean power is 45 W

To estimate the proportion of oxygen added by the desorption effect, the $TA\_OES(O)/I_{int}$ ratio was measured as a function of the injected % $O_2$, with and without additional heating (figure 7). First, for an injected % $O_2$ higher than 1.2, the heating effect seems to be negligible. Since the fully oxidized mode is reached at a % $O_2$ over 1.2, as mentioned in the section 3.1, this effect is not noticeable on the sputtering. In addition, from 0.1 % $O_2$ injected to 1.2 % $O_2$, the difference between the two curves decreases progressively, which reflects an important modification due to the heated-induced desorption of oxygen-containing molecule from the surfaces/walls. In other words, this behaviour highlights the competition between injected oxygen adsorbing to surfaces and the oxygen desorbing due to heating. Comparing the ratios at 0 % $O_2$ injected, no emission of the considered oxygen line is measured without additional heating. Therefore, the amount of oxygen released under heating leads to a similar conditions process of 0.2 % $O_2$ (highlighted by the black rectangle), from the ratio point of view.

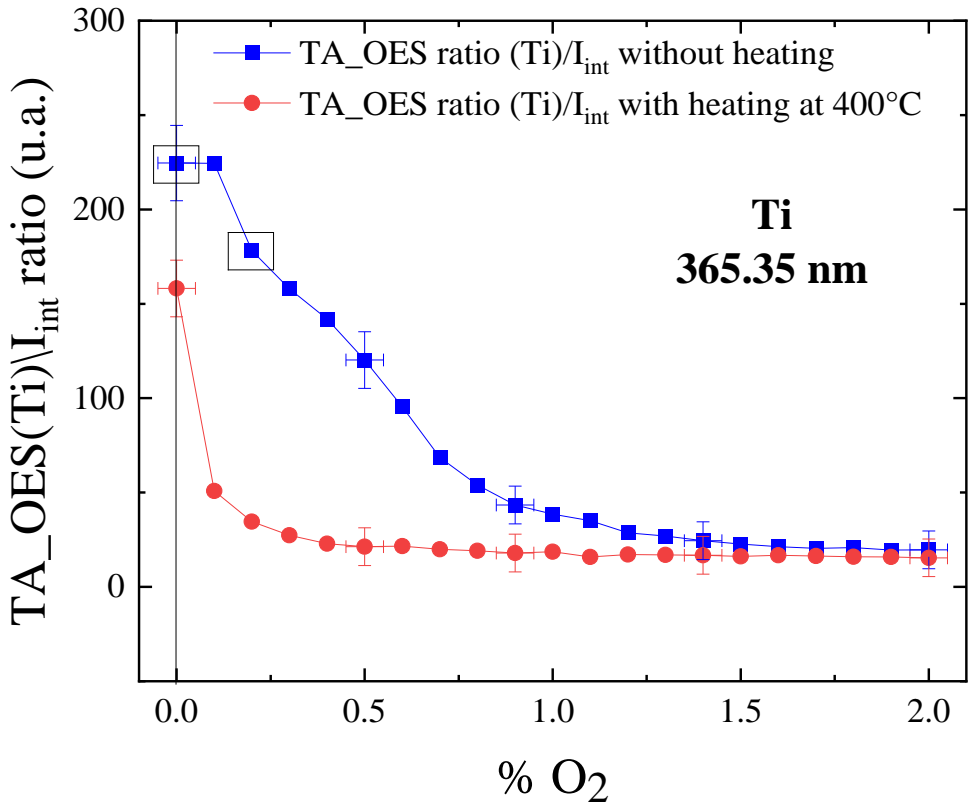

*Figure 8:* *Evolution of the ratio TA_OES(Ti)/$I_{int}$ as a function of $O_2$ percentage in the gas mixture, with heating at 400 °C and without heating. The pressure is 0.7 Pa, the HiPIMS period is 1 ms, $T_d$ = 20 µs and the mean power is 45 W*

Since our previous study [29] highlighted a good correlation between the evolution of *TA_OES(Ti)/$I_{int}$* ratio and thickness measurement as a function of the injected % $O_2$, the evolution of *TA_OES(Ti)/$I_{int}$* ratio has been investigated to corroborate the last assessment.

Figure 8 shows the evolution of *TA_OES(Ti)/$I_{int}$* ratio throughout a transition, from metallic to oxidized mode, with and without additional heating. In general, the fully oxidized mode is reached at a lower % $O_2$ (from about 1.4 % $O_2$ against 0.5) under heating (red curve). This behaviour can be attributed to the heating-induced desorption of oxygen discussed previously. By comparing the circled values on the blue curve shown in figure 8, *i.e.* at 0 % $O_2$ (≈ 225 a.u.) and 0.2 % $O_2$ (≈ 175 a.u.), the intensity ratio was ~1.3, which is the same factor obtained from the thickness ratio of unheated and heated samples done without any injected oxygen (Table 2). Furthermore, the intensity ratio obtained at 0 % $O_2$ for the heated (≈ 158 a.u.) and unheated (≈ 225 a.u.) transition is closed to 1.3, ≈ 1.4 strictly.

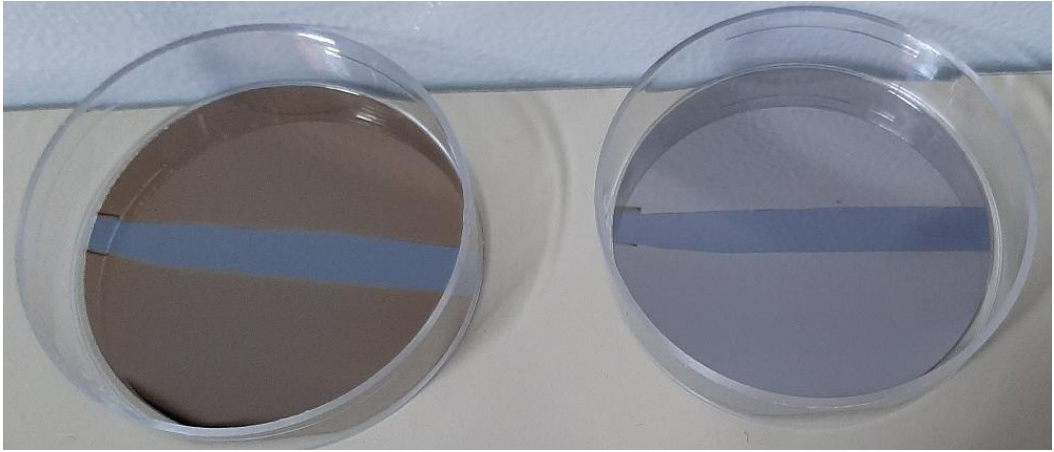

***Figure 9:*** *Photograph of two coatings made at 0 % $O_2$, with heating, coating number 10, (left) and without heating, coating number 11, (right). The pressure is 0.7 Pa, the deposition time is 4 hours, the HiPIMS period is 1 ms, $T_d$ = 20 μs and the mean power is 45 W*

Only to visualize the samples and reinforce the heating-induced desorption of the wall effect on the sputtering process, figure 9 presents a photograph of samples obtained with and without heating. Although the grey colour observed for the unheated coating (on the right) is characteristic of a metallic titanium layer, the brownish one reflects the presence of an oxide titanium in the heated one. In the middle of the substrates, the non-deposit part corresponds to the position of the cover used to obtain a step allowing the measurement of the thickness by profilometry. This observational method is obviously rudimentary but very useful to detect in first approach an eventual desorption (or leaking effect coming from the reactor) of $O_2$ during sputtering deposition process of $TiO_2$. Further material characterization (such as XRD, XPS) could have done for a better understanding of the chemical composition of the obtained coating. Nevertheless, the extensive material characterizations are not part of our study which deals with the use of the coupled optical and electrical method, presented here as a monitoring tool for sputtering processes and associated drift behaviours.

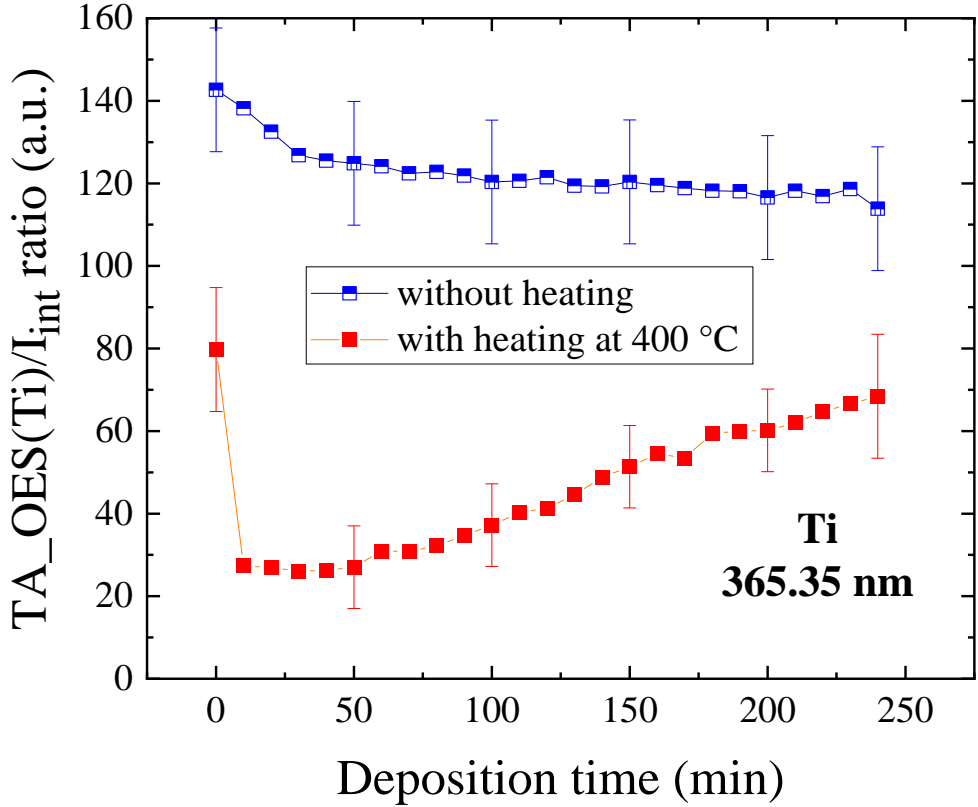

**Figure 10:** *Evolution of TA_OES(Ti)/$I_{int}$ ratio during the coating, with heating at 400 °C (red curve) and without heating (blue curve). The pressure is 0.7 Pa, the deposition time is 4 hours, the HiPIMS period is 1 ms, 0.3 %$O_2$ in the Ar/$O_2$ gas mixture, $T_d = 20$ μs and the mean power is 45 W*

To close this section, the *TA_OES(Ti)/$I_{int}$* ratio was investigated throughout a 4-hours process with and without additional heating, with 0.3 % $O_2$ injected (figure 10). The evolution of these ratios gathered in figure 10 is similar to those depicted by figure 4, where no oxygen is injected.

Whilst the heating-induced desorption effect started to decrease after 25 min of the deposition process with heating, for the 0 % $O_2$ condition (*c.f.* figure 6), the excessed oxygen from the oxygen-containing molecule outgassing of the wall seems to be non-negligible at the end of the 0.3 % $O_2$ process (figure 10, red curve). Indeed, the monotonous variation of the *TA_OES(Ti)/$I_{int}$* ratio indicates that a couple of additional hours are required to reach the unheated one (figure 10, blue curve). As the degassing lasts longer when 0.3 %$O_2$ is added to the gas mixture, a competition between the desorption and adsorption processes seems to occur, lowering the efficiency of the heating-induced desorption.

### 3.3. Quality assessment of the methodology

To test the sturdiness of the optic-electrical monitoring method presented in our previous article [29], reinforced by the latter parameters (target erosion and heating effect), a new set of experiment has been carried out using the same target and random settings, following the synopsis diagrammed in figure 1.

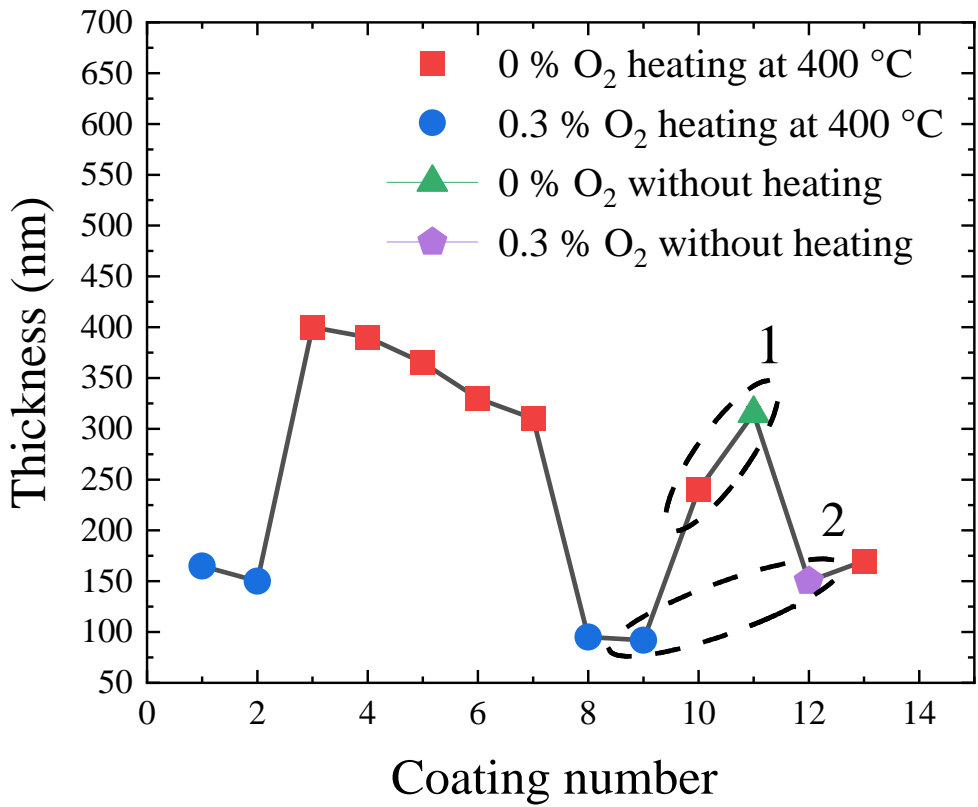

*Figure 11:* *Thickness of coatings target at different conditions. The pressure is 0.7 Pa, the deposition time is 4 hours, the HiPIMS period is 1 ms, 0 or 0.3 % $O_2$ injected in the $Ar/O_2$ gas mixture, $T_d$ = 20 µs and the mean power is 45 W. The silicon substrates are placed 11 cm in front of the target and the duration of the deposits is 4 hours. The other parameters are given in table 1.*

Figure 11 plot out the thickness of the obtained layers, which are chronologically indexed. For ease of reading, the colour code used is similar to the one reported in table 1. To start with, two main reiterated conditions are represented. The first ones are plotted in blue squares, which correspond to coatings done under 400 °C with 0.3 % $O_2$ injected in the gas mixture. The second ones are plotted in red squares, which correspond to coatings also done under 400 °C but without any oxygen injected in the gas mixture. As expected, for either of these conditions, the thickness decreases as a function of the time spent using the target, *i.e.* its erosion. Indeed, as two transitions (with and without additional heating) are measured and multiple conditioning steps of the target precede the coating process (*c.f.* figure 1), the erosion profile evolves, leading to lowering the deposition rate as discussed in section 3.1. Once the racetrack shape is well established, the thickness loss factor is less between two consecutive blue squares (*e.g.* samples n° 8 and 9) compared to that between two consecutive red squares (*e.g.* samples n° 6 and 7) since the sputtering yield is lower at 0.3 % $O_2$, than at 0 % $O_2$. In addition, the 30 % loss in thickness between samples n° 2 and 8 is due to the high sputtering yield experiments conducted between these two samples.

Furthermore, an ellipse highlights the heating effect, *ceteris paribus*. For both 0 and 0.3 % $O_2$, the deposition made without additional heating is 1.3 times thicker than the one made under 400 °C. These results support the heating-induced desorption effect discussed previously where the oxygen outgassing the heated wall tends to lower the deposition rate and the sputtering yield by oxidation of the target.

Finally, the *TA_OES(Ti)/$I_{int}$* ratio was measured every 10 min during each coating experiment listed in table 1. To compare it with the variation of the final thickness, the *TA_OES(Ti)/$I_{int}$* values have been averaged over the 4 hours, noted *<TA_OES(Ti)/$I_{int}$>* ratio and plotted in figure 12.

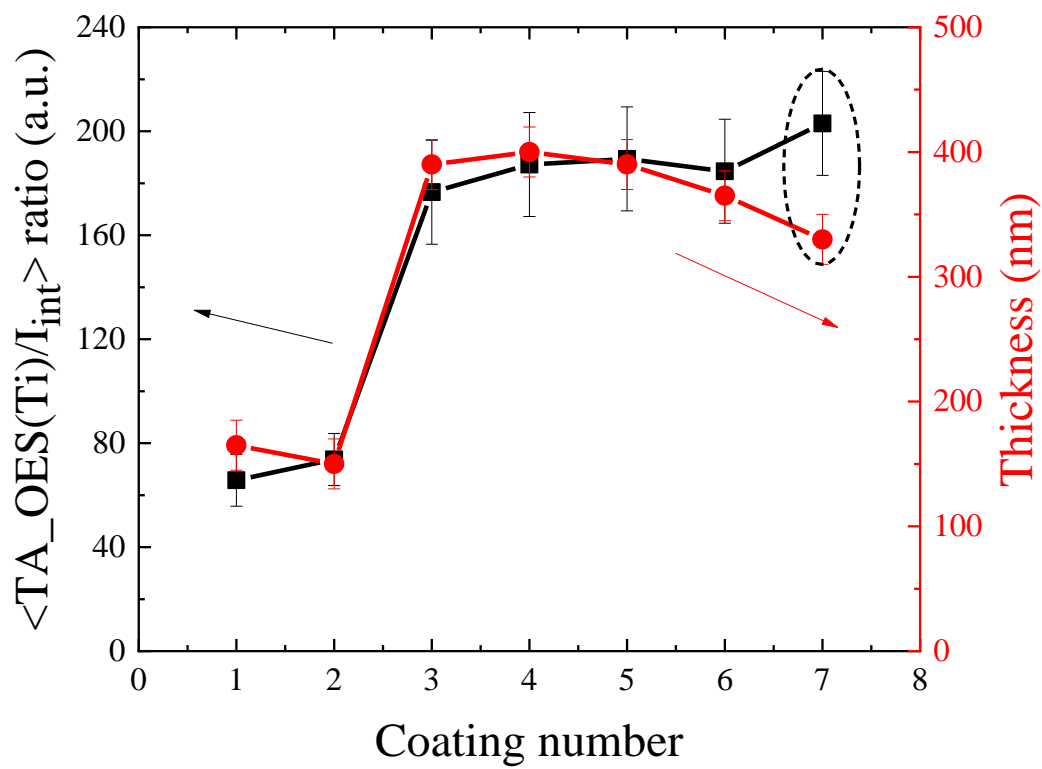
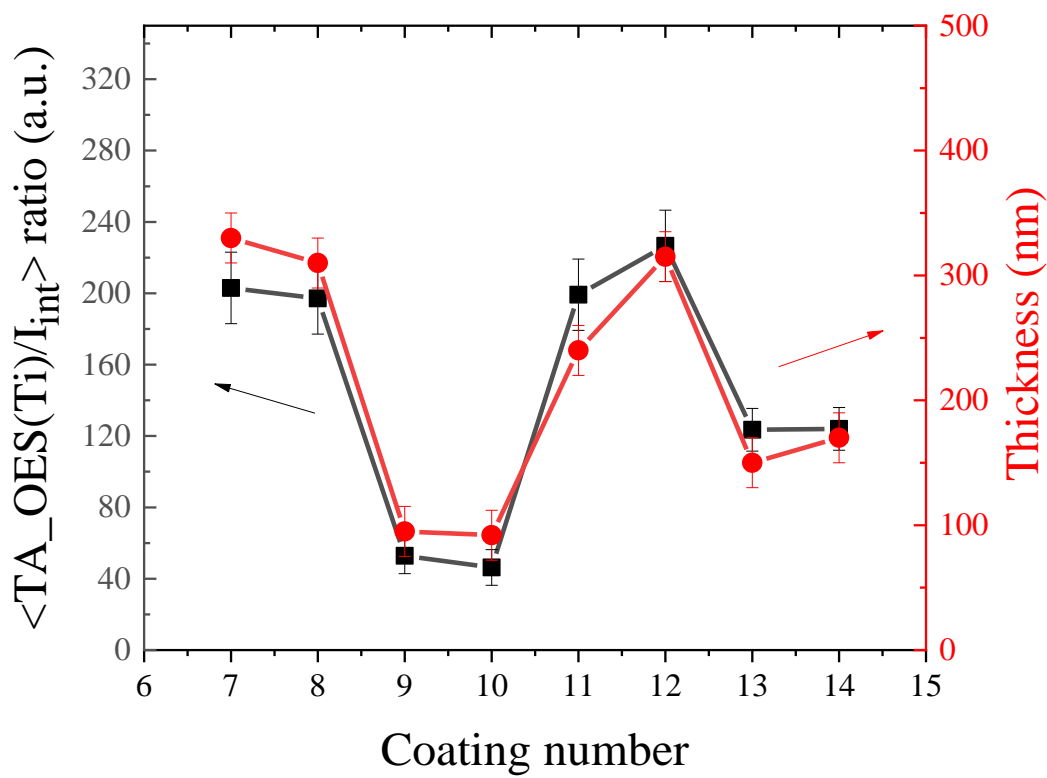

*Figure 12:* Comparison between the thickness of deposits and the corresponding $<TA\_OES(Ti)/I_{int}>$ mean ratio. The pressure is 0.7 Pa, the HiPIMS period is 1 ms, $T_d$ = 20 µs

*and the mean power 45 W. The deposition time is 4h and the silicon substrates are placed 11 cm away from the target. The other conditions are shown in table 1.*

Figure 12 exhibits a good accordance between these two quantities, for the samples from n° 1 to 5. Concerning sample n° 6, the 15 – 20 % difference is explained by a misalignment of the fibre subsequent to the loan of the spectrometer, which is a shared device. To take this modification into account, the $<TA\_SOE(Ti)/I_{int}>$ mean ratio y-axis (at the left of the plot) of the following samples (n° 6 to 13) was adjusted, keeping the thickness y-axis (at the right of the plot) the same. Even though, the evolutions of $<TA\_SOE(Ti)/I_{int}>$ *average* ratio and thickness still agree relatively well especially since the conditions have been, voluntarily and radically, changed.

These last results, based on the optical and electrical measurements done during the coating process, seem to demonstrate the feasibility to track and finely monitor the deposition process "continuously", at least with the Titanium-Argon-Oxygen system under the conditions previously studied. Since the time spent to do a measure (20 s in our case, time spent to average out the optical and electrical measurements) is much shorter than the deposition time (4 h in our case), the monitoring of the process can be estimated as a "real time" control of the thickness.

## Conclusion:

In this study, a method is proposed to enforce the feedback control loop on electrical parameters with coupled electrical and optical emission spectroscopy measurements by including process drifts. First, the erosion profile study brought out a decrease in the deposition rate depending on the racetrack shape, linked to the time spent using a same target. Indeed, the $TA\_OES(Ti)/I_{int}$ ratio and the coating thickness for a new target, which has a flat surface, is ≈ 2.5 times higher than a few hundred hours used target in our conditions, with a final erosion depth of about 1.6 mm. This behaviour can be explained with the shape of the sputtered surface. With a new target, the sputtered atoms are mainly ejected perpendicularly to the initial surface. As soon as the target is eroded, a part of these sputtered atoms caries a radial velocity component (parallel to the initial target surface) higher and higher with the erosion state. Therefore, these atoms may be more and more deflected towards the walls and not detected by the emission measurement since they no longer pass through the densest HiPIMS plasma centred above or in front of the racetrack.

Second, an additional substrate-holder heating of 400 °C has been found to heat the reactor walls and induce oxygen outgassing from them, in our conditions. The amount of desorbed oxygen has been estimated at an equivalent gas mixture of 0.2 % $O_2$, whereas only pure argon is injected, after each venting. Consequently, working at low oxygen injected ratio (≤ 1 % $O_2$), the heating-induced desorption is non-negligible through noticeable with $TA\_OES(Ti)/I_{int}$ ratio and thickness measurement which are around 30 % lower than the unheated deposition process. During the 4-hours process under pure argon with additional heating, the temporal optical and electrical measurements highlighted an increase of the oxygen emission line (at 777 nm) the first dozens minutes before decreasing. At the same time, the titanium emission line (at 365 nm) evolves in an opposite way since the amount of oxygen implies the sputtering yield.

Implementing this coupled optical and electrical method into a software could allow to approach real-time monitoring of the deposition process as the period measurement is shorten with automation. Indeed, the most critical time is the one spends to measure and process data from the $TA\_OES/I_{int}$ ratio. From this, a set of coatings, with a constant thickness value could be realized by tuning "in real-time" the applied HiPIMS power or correcting the duration of the process depending on the variation of $TA\_OES/I_{int}$ ratio since the target erosion and the wall conditions are taken into account. If tuning the applied power implies non-negligible modifications of the target and the coating oxidation state, one could also couple this method with those mainly developed so far [26, 32 – 35], *i.e.* feedback loops to adjust the injected reactive gas percentage.

## Acknowledge:

This work was supported by the European Regional Development Fund through the Interreg Grande Region-PULSATEC project.